# Enhancing Supply Chain Resilience Through Metaverse and ChatGPT Technologies

**Sarhir Oumaima, Dr. Zoubida BENMAMOUN, Dr. Mouad BEN MAMOUN**

## Abstract :

Global supply lines have been severely disrupted by the COVID-19 epidemic and the conflict between Russia and Ukraine, which has sharply increased the price of commodities and generated inflation. These incidents highlight how critical it is to improve supply chain resilience (SCRES) in order to fend off unforeseen setbacks. Controlling both internal and external interruptions, such as transportation problems brought on by natural catastrophes and wars, is the responsibility of SCRES.
Enhancing resilience in supply chains requires accurate and timely information transfer.

Promising answers to these problems can be found in the Metaverse and ChatGPT, two new digital technologies. The Metaverse may imitate real-world situations and offer dynamic, real-time 3D representations of supply chain data by integrating blockchain, IoT, network connection, and computer power.Large-scale natural language processing model ChatGPT improves communication and data translation accuracy and speed.
To manage risk and facilitate decision making in Supply Chain management, firms should increase information transmission, Speed and quality.
This study aim to show the importance of ChatGPT and Metaverse technologies to improve SCRES, with an emphasis on the most important criteria for SCRES, and maturity factor that can influence directly the SC development.

## 1. Introduction:

Global supply chains have experienced unprecedented strain the past few years. Due to the covid pandemic and Russia's invasion of Ukraine, impacted supply chains cause skyrocketing commodity prices and inflation. Such incidents have exposed supply chain network weaknesses and emphasized the need to improve SC resilience (SCRES). What supplies chain resilience elements refer to the ability of a supply chain to plan for its response to withstand and recover from any surprise disruption?

In the face of these issues, the strategic incorporation of sophisticated digital technologies has become one of the key methods for enhancing supply chain resilience. Among these technologies, the Metaverse and ChatGPT can revolutionize the supply chain. The Metaverse refers to a collective virtual environment, generated by the merging of virtually improved physical reality and a consistently existing virtual space, including digital technologies like the Internet of Things (IoT), network communication, computing capacity, and blockchain. Then, this would lead us to a virtual environment that is very authentic and captures real-world supply chain settings. In other words, this virtual environment becomes dynamic and interactive as a simulated environment within which real-world decisions can be made in real time.

Conversely, OpenAI uses ChatGPT, a large natural language processing model, which utilizes deep learning to read and write like humans. The technology can greatly increase the communication accuracy and efficiency within supply chains, which will then support the decision processes better and automate routine processes.

There should be a paper out there that brings Metaverse and the ChatGPT thing together in combination with supply chain resilience. We hope to illustrate their potential to mend current holes in supply chains by studying them in the context of modeling supply chain environments, accelerating real-time data processing, and enabling intelligent communication. We will also mention difficulties and issues in implementing them, such as the security of the data and its integration into existing systems.

Thus, the Metaverse and ChatGPT technologies battle in the supply chain area can be a step toward a more secured, lean, and robust supply chain that can withstand the chaos in this world.

## 2. Literature Review:

1. **Supply Chain Resilience: Dimensions, Pathways, and Influencing Factors:**

Overall, these studies highlight the fact that supply chain resilience is multifaceted, with key dimensions being organizational capacity, collaboration, flexibility, and information systems (Michel, Gerbaix, and Marc, 2023).

These are developed by digital transformation, supply chain diversification, and supplier centralization, etc. (Weili Yin, Wenxue Ran, 2022).

There are also six critical resilience elements -- product supply, resources, partners, information response, capital, and knowledge -- that are suggested to highly impact resilience (Xianjun Zhu, Yenchun Jim Wu, 2023).

2. **Impact of Emerging Technologies on Supply Chain Management:**

Emerging technologies like the metaverse, AI (ChatGPT), and digital twins are transforming supply chain management. The metaverse enhances traceability and sales in agri-food supply chains and fosters green knowledge sharing and rational decision-making (Asmae El Jaouhari, Jabir Arif, Fouad Jawab, Ashutosh Samadhiya, Anil Kumar, 2024 & Ping-Kuo Chen, Xiang Huang, 2023). Platforms like NVIDIA Omniverse enable real-time simulation and planning, boosting resilience through digital twins (Dmitry Ivanov, Jennifer Blackhurst, Ajay Das, 2023). ChatGPT offers capabilities in route optimization, predictive maintenance, and data analysis, significantly improving operational efficiencies and decision-making in supply chains (Guilherme Francisco Frederico, 2023).

3. **Technological Disruptions and Their Broader Implications:**

Disruptive technologies such as blockchain, DAOs, and AI are reshaping not only supply chains but also broader networks like e-diasporas (Igor Calzada, 2023). The Industrial Metaverse, integrating extended reality and next-gen communication networks, shows potential for enhancing industrial production, despite challenges related to security and interoperability (Shiying Zhang, Jun Li, Long Shi, Ming Ding, Dinh C. Nguyen, Wen Chen, Zhu Han, 2024). The adoption of Gen-AI/ChatGPT in operations and supply chain management presents both significant benefits in efficiency and performance, and challenges related to security and ethical concerns (Alexandre Dolgui, Dmitry Ivanov, 2023).

| Capabilities | Description | Practices | References |
|---|---|---|---|
| C1: Organizational Capacity | The ability of an organization to manage resources effectively and adapt to changing circumstances is crucial for resilience. | - Training and Development.<br>- Resource Allocation.<br>- Performance Metrics. | *Sylvie Michel, Sylvie Gerbaix, Bernard Marc 2023; Maureen S. Golan, Laura H. Jernegan, Igor Linkov 2020; Samuel Fosso Wamba, Maciel M. Queiroz, Charbel Jose Chiappetta Jabbour, Chunming (Victor) Shi 2023;* |
| C2: Collaboration | Strong partnerships and collaboration among stakeholders enhance information sharing and resource allocation during disruptions. The Metaverse provides 3D virtual collaboration environments that enhance supply chain visibility and resilience. | - Stakeholder Engagement.<br>- Joint Planning Information.<br>- Sharing Platforms. | *Weili Yin, Wenxue Ran 2022; Sylvie Michel, Sylvie Gerbaix, Bernard Marc 2023; Kirstin Scholten, Sanne S childer 2015;* |
| C3: Flexibility | The capacity to adjust operations, processes, and supply chain strategies in response to unexpected changes is essential for maintaining resilience. | - Agile Processes.<br>- Multi-Sourcing Strategies.<br>- Scenario Planning.<br>- Supplier Network Expansion. | *Xianjun Zhu, Yenchun Jim Wu 2022; Sylvie Michel, Sylvie Gerbaix, Bernard Marc 2023; Benjamin R. Tukamuhabwa, Mark Stevenson, Jerry Busby, Marta Zorzini 2015;* |
| C4: Sustainability | Integrating sustainability into supply chain operations can enhance long-term resilience by addressing environmental and social risks. | - Sustainable Sourcing Policies.<br>- Sustainability Reporting. | *Ping-Kuo Chen, Xiang Huang 2023; Xianjun Zhu, Yenchun Jim Wu 2022; Asmae El Jaouhari, Jabir Arif, Fouad Jawab, Ashutosh Samadhiya, Anil Kumar 2024;* |
| C5: Agility | The integration of Metaverse and ChatGPT technologies promotes agility by enabling dynamic response capabilities in changing market conditions. | - Agility in Operations.<br>- Supplier Diversification. | *Benjamin R. Tukamuhabwa, Mark Stevenson, Jerry Busby, Marta Zorzini 2015; Dmitry Ivanov, Jennifer Blackhurst, Ajay Das 2021;* |
| C6: Visibility | The ability to track and monitor the flow of goods, information, and resources throughout the supply chain in real-time. | - Predictive Analytics.<br>- Real-Time Tracking.<br>- Demand Forecasting. | *Alexandre Dolgui, Dmitry Ivanov 2023; Dmitry Ivanov, Jennifer Blackhurst, Ajay Das 2021;* |

| | | | *Kirstin Scholten, Sanne S childer 2015;* |
|---|---|---|---|
| C7: Cost Optimization | The use of ChatGPT in supply chains can lead to significant cost reductions by automating administrative tasks and improving overall performance. | <ul><li>Data Analytics for Optimization.</li><li>Lean Inventory Management.</li><li>Performance Metrics Monitoring.</li></ul> | *Rohit Raj, Arpit Singh, Vimal Kumar, Pratima Verma 2023; Guilherme Francisco Frederico 2023;* |
| C8: Security | The Metaverse enhances supply chain security through technologies like blockchain for traceability and transparency | <ul><li>Cybersecurity Measures.</li><li>Blockchain. Supply Chain</li><li>Risk Assessment.</li></ul> | *Rohit Raj, Arpit Singh, Vimal Kumar, Pratima Verma 2023; Abid Haleem, Mohd Javaid, Ravi Pratap Singh 2022; Stephen Rice. Sean R Crouse, Scott R. Winter, Connor Rice 2024; Shiying Zhang, Jun Li, Long Shi, Ming Ding, Dinh C. Nguyen, Wen Chen, Zhu Han 2024;* |
| C9: Technological integration | Implementing digital technologies and Advanced information systems facilitate real-time data sharing and decision-making, which are critical during emergencies. | <ul><li>Digital Twins Collaboration Platforms.</li><li>Automation of processes.</li><li>Implementation of IoT Solutions.</li></ul> | *Dmitry Ivanov, Jennifer Blackhurst, Ajay Das 2021; Weili Yin, Wenxue Ran 2022; Sylvie Michel, Sylvie Gerbaix, Bernard Marc 2023;* |
| C10: Risk Prevention Awareness | To develop a proactive approch to identify and mitigate risks before they became a crises. | <ul><li>Risk Identification.</li><li>Regular Risk Audits.</li><li>Proactive Risk Mitigation Plans.</li></ul> | *Xianjun Zhu, Yenchun Jim Wu 2022; Youan Ke, Lin Lu, Xiaochun Luo 2023; Dmitry Ivanov, Jennifer Blackhurst, Ajay Das 2021; Samuel Fosso Wamba, Maciel M. Queiroz, Charbel Jose Chiappetta Jabbour, Chunming (Victor) Shi 2023;* |

**Table 1.** Capabilities of Supply Chain Resilience

The strategic utilization of digital transformation, artificial intelligence (AI), and metaverse technologies is becoming a pivotal factor in strengthening the supply chain resilience, which allows businesses to overcome disturbances and continue their operations. By continually aligning these sophisticated technologies within ever-changing market dynamics, specific operating circumstances, and robustness routes, a company can find a way to ensure long-term success. With digital transformation, data can be analyzed in real time, with the use of predictive modeling and automation, enabling the organization to predict the risk as well as optimize its decision-making process. Likewise, artificial intelligence facilitates supply chain visibility, predicts variations in demand, and optimizes the logistics process; metaverse technologies implement virtual immersive environments that aid in collaboration, training, and remote operations. (Oyku_Ilgar 2023)

The humanitarian values and green supply chain practices can also help improve sustainable supply chain management besides technological advancement. The concept of "Greening" emphasizes the importance of sharing knowledge and implementing strategies that contribute to environmentally friendly practices. Virtual environments and AI reinforce these initiatives by allowing resources to be allocated efficiently, waste to be minimized, or energy to be optimized. Yet, the embrace of such a revamping technology as this does not come easily. There are evolving risks associated with cybersecurity, information security, ethics, and exploitable weaknesses in operations. A vigilant approach to risk management and continuous learning within the organization become necessities. However, to address these challenges, businesses will need to establish robust governance structures, spend resources on employee education, and foster an atmosphere of creativity and flexibility. (M. Ali Ülkü, ORCID, James H. Bookbinder, and Nam Yi Yun, 2023)

Empirical and qualitative research provides valuable insights into the practical applications of these resilience strategies, offering case studies and real-world examples that guide managers in decision-making during crises. Such studies highlight the benefits of leveraging technology to build adaptive, responsive supply chains while also revealing potential pitfalls and best practices for mitigating associated risks. Ultimately, the intersection of digital transformation, AI, metaverse technologies, and sustainable, human-centric practices forms the foundation of resilient, future-proof supply chains. This holistic approach not only safeguards organizations against unforeseen disruptions but also positions them as leaders in innovation, sustainability, and operational excellence. (Meriem Riad 1, Mohamed Naimi, and Chafik Okar 2024)

# 3. Methodology:

To effectively enhance supply chain resilience using new techmologies, we will adopt a two-phase approach that leverages the strengths of both the Analytic Hierarchy Process (AHP) and Grey-TOPSIS (G-TOPSIS) methodologies.

In the first phase, AHP will be employed to prioritize a set of critical organizational capabilities, including Visibility, Agility, Flexibility, Security, Sustainability, Collaboration, Technology Integration, Risk Prevention Awareness, and Organizational Capacity. By decomposing these capabilities into a hierarchical structure, AHP enables decision-makers to conduct systematic pairwise comparisons, allowing for the quantification of subjective judgments and the assignment of relative weights to each capability. Such an organized process not only helps to clarify the most compelling factors but also keeps important objectives in check. AHP (Analytic Hierarchy Process) promotes buy-in from all stakeholders by making the process both qualitative and quantitative. Prioritization can be extremely subjective, and inexperienced individuals often base decisions on seemingly objective factors such as severity or age of the problem. (Vargas 2010, opp. 12-13)

In the second phase, the Grey-TOPSIS (G-TOPSIS) methodology will be applied to identify and rank high-maturity organizational practices that contribute to the enhancement and development of the prioritized capabilities. G-TOPSIS, an extension of the classical TOPSIS method, integrates Grey System Theory to address the challenges posed by incomplete, uncertain, or imprecise data—common issues in dynamic and complex organizational environments. The technique assesses each practice relative to the ideal solution based on proximity: in this way, the uncertainty of decision-making situations can clearly be expressed. G-TOPSIS allows organizations to translate their theoretical prioritization into a practical, applicable form and focus on "best bang for your buck" kinds of practices; thus ensuring that what resources they do have allocated towards capability development are used in the most efficient manner. It ensures a comprehensive evaluation process that enables organizations to improve their capabilities, reduce risks, and quickly and resiliently respond to changing market demands. (Khadija Echefaj, Abdelkabir Charkaoui, Anass Cherrafi, Anil Kumar, Sunil Luthra 2022)

## 1. AHP Methodology:

The Analytic Hierarchy Process (AHP) is a prominent method within Multi-Criteria Decision-Making (MCDM), a field dedicated to evaluating and prioritizing options based on multiple, often conflicting criteria. Developed by Thomas Saaty in the 1970s, AHP assists decision-makers in structuring complex problems into a hierarchical framework, facilitating pairwise comparisons among criteria and alternatives. (Hamed Taherdoost and Mitra Madanchian 2023 ) This structured approach enables the assignment of numerical weights to various factors, reflecting their relative importance, and supports the synthesis of these weights to determine the most suitable decision option. AHP's systematic methodology has been widely applied across diverse fields, including finance, engineering design, and sustainable development, due to its capacity to handle both qualitative and quantitative data effectively. By decomposing decision problems into a hierarchy of subproblems, AHP allows for a comprehensive analysis that incorporates both objective data and subjective judgments, enhancing the robustness and transparency of the decision-making process. (Nitin Rane, Anand Achari, and Saurabh Purushottam Choudhary 2023)

To create a framework using the Analytic Hierarchy Process (AHP) methodology, considering the criteria identified in the context of supply chain resilience, follow these structured steps:

**Step 1: Define the Problem and Goal**

Clearly articulate the objective of your AHP analysis. For instance, the goal could be to evaluate and prioritize dimensions of supply chain resilience such as organizational capacity, collaboration, flexibility, and humanitarian culture.

**Step 2: Identify Criteria**

Based on the finding from the literature review, list the main criteria for supply chain resilience:

- **Organizational Capacity**
- **Collaboration**
- **Flexibility**
- **Sustainability**
- **Agility**
- **Visibility**
- **Cost Optimization**
- **Security**
- **Technological Integration**

- **Risk Prevention Awareness**

**Step 3: Pairwise Comparisons**

Conduct pairwise comparisons for each criterion to determine their relative importance. Use a scale (e.g., 1 to 9) where:

- **1 indicates equal importance,**
- **3 indicates moderate importance,**
- **5 indicates strong importance,**
- **7 indicates very strong importance,**
- **9 indicates extreme importance.**

Create a comparison matrix for each level of criteria.

**Step 4: Weights Calculation**

Using the pairwise comparison matrices, calculate the weights for each criterion.

The process begins by extracting the pairwise comparison matrix, referred to as matrix A, from the data gathered during the interviews. The principal right eigenvector of matrix A is then calculated and denoted as 'w'.

If the matrix is incompatible and in case of incomplete consistency, pair comparisons matrix cannot be used normalizing column to get Wi. (Jalaliyoon, et al., 2012).

For a positive and reversed matrix, Eigenvector technique can be used which in it:

$$e^T(1,1,\cdots,1)$$

the following formula is applied to transform the raw data into meaningful absolute values and normalized weight w = (w1, w2, w3… wn).

Aw = λmax w, λmax ≥ n

$$\Lambda\max = \frac{\sum a_j w_j - n}{w_1}$$

A={aij} with aij=1/ aij

A: pair wise comparison

w: normalized weight vector

λmax : maximum eigen value of matrix A

aij: numerical comparison between the values i and j

**Step 5: Consistency Evaluation**

In this step, to validate the AHP results, the consistency ratio (CR) is calculated using the formula:

$$CR = CI/RI.$$

Here, the consistency index (CI) is determined through the application of the following formula:

$$CI = \frac{\lambda_1 \max - n}{n - 1}$$

Where the value of RI is extracted fron the following Table (Hamed Taherdoost 2017)

| Dimension | RI |
|---|---|
| 1 | 0 |
| 2 | 0 |
| 3 | 0.5799 |
| 4 | 0.8921 |
| 5 | 1.1159 |
| 6 | 1.2358 |
| 7 | 1.3322 |
| 8 | 1.3952 |
| 9 | 1.4537 |
| 10 | 1.4882 |

**Table 2.** The Value of Random Consistency Index

## 2. Grey TOPSIS:

G-TOPSIS is an extension of the Technique for Order of Preference by Similarity to Ideal Solution (TOPSIS), a widely used multi-criteria decision-making (MCDM) method. Traditional TOPSIS evaluates alternatives based on their geometric distance from an ideal solution, identifying the option closest to the ideal and farthest from the negative-ideal. G-TOPSIS improves on this method by applying Grey System Theory, which is capable of dealing with uncertainty and information that is missing. (Re. K., Sohn, J., & Liu, X. 2010) This combination enables the G-TOPSIS to handle cases when data are missing or vague and is well suited for complicated decision-making systems. For example, G-TOPSIS has been used in the selection of cloud services, in which decision-makers have to potentially select one of numerous providers against incomplete or uncertain criteria. By integrating grey relational analysis into the traditional TOPSIS, G-TOPSIS offers a more comprehensive strategy for the ranking and selection of alternatives when faced with uncertainty, thus enhancing the validity of the decision-making process. Richard Nyaogaa, Peterson Magutu b, and Mingzheng Wang c, 2016).

**Grey TOPSIS Methodology Steps:**

- **Define the Decision Matrix**: Construct a decision matrix that includes the alternatives and criteria. Each element represents the performance of an alternative on a specific criterion.
- **Normalize the Decision Matrix**: Convert the decision matrix into a normalized form to make different units comparable. This often involves using vector normalization techniques.
- **Determine the Weighted Normalized Decision Matrix**: Assign weights to each criterion based on their importance and multiply the normalized values by these weights.
- **Identify the Ideal and Anti-Ideal Solutions**: Determine the best (ideal) and worst (anti-ideal) values for each criterion.

- **Calculate the Separation Measures**: Compute the separation of each alternative from the ideal and anti-ideal solutions using a distance measure.
- **Compute the Relative Closeness Coefficient**: Calculate the closeness of each alternative to the ideal solution, allowing for the ranking of alternatives. (Yang, J.B., & Xu, D.L. 2016)

# 4. Results and finding:

This study seeks to identify essential capabilities and high maturity factors that enhance supply chain resilience. The findings are organized into two sub-sections for clarity. The first sub-section presents a ranking of the identified capabilities due to AHP methodology, while the second sub-section discusses the results obtained from the G-TOPSIS analysis.

## 1. AHP Application:

The process commenced with the creation of a pairwise comparison matrix to determine the individual weight scores of the experts. This study employed a multi-criteria decision-making approach to prioritize capabilities essential for a resilient supply chain. Drawing from the literature review, we identified eleven key supply chain capabilities: flexibility, collaboration, visibility, security, agility, organizational capacity, cost optimization, sustainability, technological integration, and risk prevention awareness. In the first stage of the Analytic Hierarchy Process (AHP), we calculated the weights for these ten capabilities, which were further refined in the second stage.

We asked experts to rank the 10 criteria in order of priority though an online interview. The opinion of 6 experts was taken into consideration.

## 2. Capacities ranking:

The table below shows the rank and the weight of resilience capabilities. And sustainability is given (13.5%), which is the most important for supply chain sustainability. Agility follows with 11.9%, underscoring the need for supply chains to be highly responsive to changing market conditions and disruptions in the marketplace. This ability enables companies to react quickly to changes, providing the adaptability needed to remain strong. Cost (10.9%) This ranked third. The significance of this lies in the company's realization that cost containment is crucial, as escalating expenses lower the profit margin unless mitigated by improved service levels. Managing costs effectively leads to more resilience, as firms are better able to survive when financial pressures increase in times of crisis. Visibility (10.8%) - Comprehensive insight into the supply chain enables better comprehension of supply chain trends and potential risks. Coming in at number four, this means that companies must spend money on technology and processes that create more supply chain visibility. And then by Technology (10.5 %) The incorporation of technology to facilitate operations and communication between the supply chain members is of less but still important concern, thus rating it fifth. Flexibility, at 10.41%, came in 6th, and indicates a requirement for supply chains to adjust their operations and frameworks in accordance with fluctuating demands and circumstances. Security (10.40%): Security is extremely important to prevent supply chains from being affected by outside influences. Risk prevention awareness (10.3%) comes in at number eight. This implies that organizations need to develop a thinking process of spotting risks, fixing the risks in order to prevent them from becoming major problems. Collaboration (5.7%) with supply chain partners is ranked ninth and indicates that although it is desirable in terms of resilience, it appears to be of less importance than most of the capabilities identified in this study. Lastly, organizational capacity (5,5%) has the lowest rating of the ten categories evaluated. Meaning that although it is still an essential part of resiliency, it can be superseded in short-term strategic planning. In a broad sense, these rankings underscore a strategic prioritization of sustainability, agility, and cost optimization as basic pillars to resilient supply chains and emphasize visibility and technological capability as supportive elements for these capabilities.

| Capabilities | Expert 1 | Expert 2 | Expert 3 | Expert 4 | Expert 5 | Expert 6 | Weight | Rank |
|---|---|---|---|---|---|---|---|---|
| Visibility | 0,27253 | 0,25836 | 0,03151 | 0,06355 | 0,01 | 0,02 | 0,10916 | 4 |
| Agility | 0,19089 | 0,18515 | 0,03877 | 0,04681 | 0,11 | 0,14 | 0,11897 | 2 |
| Collaboration | 0,10664 | 0,10749 | 0,05912 | 0,04135 | 0,01 | 0,02 | 0,05743 | 9 |
| Flexibility | 0,10602 | 0,09493 | 0,04497 | 0,04252 | 0,29 | 0,05 | 0,10402 | 6 |
| Security | 0,08253 | 0,05688 | 0,16852 | 0,10340 | 0,07 | 0,15 | 0,10363 | 7 |
| Cost Optimization | 0,04953 | 0,03293 | 0,09445 | 0,32961 | 0,08 | 0,08 | 0,10982 | 3 |
| Technological integration | 0,07336 | 0,15490 | 0,18116 | 0,07205 | 0,12 | 0,03 | 0,10529 | 5 |
| Risk Prevention Awareness | 0,06309 | 0,05518 | 0,18762 | 0,15184 | 0,08 | 0,08 | 0,10353 | 8 |
| Sustainability | 0,01727 | 0,01385 | 0,06405 | 0,09992 | 0,21 | 0,41 | 0,13480 | 1 |
| Organizational Capacity | 0,03814 | 0,04034 | 0,12983 | 0,04895 | 0,04 | 0,03 | 0,05429 | 10 |
| CR | 0,0204738 | 0,092675 | 0.081710 | 0,08516 | 0,095902 | 0,0275 | | |

**Table 3.** Capabilities Ranking

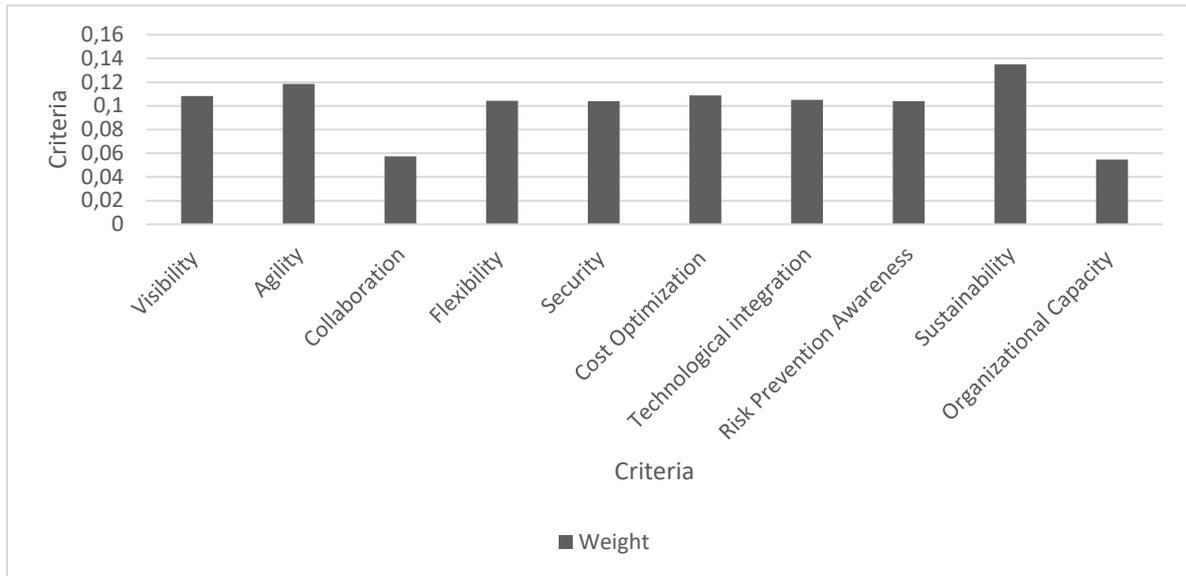

**Figure 1:** Ranking of Capabilities

### 3. G-TOPSIS results:

After ranking and defining the important capabilities to improve supply chain resilience through the new technologies. We defined 5 maturity factors that influence the development of Supply Chain.

We will apply the Grey-TOPSIS (G-TOPSIS) methodology to rank four critical maturity factors that influence the enhancement of supply chain resilience. The factors under consideration are **Technological Proficiency, Collaborative Ecosystem, Agility and Responsiveness,** and **Cost Efficiency and Optimization.** These maturity factors will be evaluated against the five more importance criteria ranked by AHP Method: **Sustainability, Agility, Cost Optimization, Visibility,** and **Technological Integration.** G-TOPSIS, by incorporating Grey System Theory, allows for effective handling of uncertainty and incomplete data, ensuring a more accurate and robust decision-making process. Through pairwise comparisons and distance-based evaluations, G-TOPSIS will enable us to identify the maturity factor that most closely aligns with the ideal solution while remaining farthest from the negative-ideal solution. This approach will provide valuable insights into which organizational practices should be prioritized to drive supply chain resilience and achieve long-term competitive advantage. (Mohammadreza SADEGHI1∗, Seyed Hossein RAZAVI1, Narges SABERI2)

TOPSIS as one of MCDM methods considers both the distance of each alternative from the positive ideal and the distance of each alternative from the negative ideal point. In other words, the best alternative should have the shortest distance from the positive ideal solution (PIS) and the longest distance from the negative ideal.

In this study there we choose the 5 best criterias and 4 alternatives as maturity factors that are ranked based on TOPSIS method. The following table shows the decision matrix.

|                                       | Sustainability   | Agility          | Cost Optimization | Visibility       | Technological integration |
|---------------------------------------|------------------|------------------|-------------------|------------------|---------------------------|
| Technological Proficiency             | 6                | 8.33333333333333 | 8                 | 7.66666666666667 | 8.66666666666667          |
| Collaborative Ecosystem               | 8.33333333333333 | 3.33333333333333 | 4                 | 6                | 3                         |
| Agility and Responsiveness            | 5.33333333333333 | 8                | 6                 | 7                | 7.66666666666667          |
| Cost Efficiency and Optimization      | 8.66666666666667 | 5                | 7.33333333333333  | 5.33333333333333 | 5                         |

**Table 4.** Grey Topsis Decision Matrix

**STEP 1: Normalize the decision-matrix.**

The following formula can be used to normalize.

$$r_{ij}(\text{x}) = \frac{x_{ij}}{\sqrt{\sum_{i=1}^{m} x_{ij}^2}} \quad i = 1, \dots, m \ ; j = 1, \dots, n$$

The following table shows the normalized matrix.

|                                  | Sustainability | Agility | Cost Optimization | Visibility | Technological integration |
|----------------------------------|----------------|---------|-------------------|------------|---------------------------|
| Technological Proficiency        | 0.415          | 0.64    | 0.614             | 0.584      | 0.669                     |
| Collaborative Ecosystem          | 0.576          | 0.256   | 0.307             | 0.457      | 0.232                     |
| Agility and Responsiveness       | 0.369          | 0.614   | 0.46              | 0.533      | 0.592                     |
| Cost Efficiency and Optimization | 0.599          | 0.384   | 0.563             | 0.406      | 0.386                     |

**Table 5.** The normalized matrix

**STEP 2: Calculate the weighted normalized decision matrix.**

According to the following formula, the normalized matrix is multiplied by the weight of the criteria.

$v_{ij}(x) = w_j r_{ij}(x) \quad i = 1, \ldots, m \; ; j = 1, \ldots, n$

The following table shows the weighted normalized decision matrix.

|  | Sustainability | Agility | Cost Optimization | Visibility | Technological integration |
|---|---|---|---|---|---|
| Technological Proficiency | 0.083 | 0.128 | 0.123 | 0.117 | 0.134 |
| Collaborative Ecosystem | 0.115 | 0.051 | 0.061 | 0.091 | 0.046 |
| Agility and Responsiveness | 0.074 | 0.123 | 0.092 | 0.107 | 0.118 |
| Cost Efficiency and Optimization | 0.12 | 0.077 | 0.113 | 0.081 | 0.077 |

**Table 6.** The weighted normalized matrix

**STEP 3: Determine the positive ideal and negative ideal solutions.**

The aim of the TOPSIS method is to calculate the degree of distance of each alternative from positive and negative ideals. Therefore, in this step, the positive and negative ideal solutions are determined according to the following formulas.

$A^+ = (v_1^+, v_2^+, \ldots, v_n^+)$

$A^- = (v_1^-, v_2^-, \ldots, v_n^{-+})$

So that

$v_j^+ = \{(\max v_{ij}(x) | j \epsilon j_1), (\min v_{ij}(x) | j \epsilon j_2)\} \; i = 1, \ldots, m$

$v_j^- = \{(\min v_{ij}(x) | j \epsilon j_1), (\max v_{ij}(x) | j \epsilon j_2)\} \; i = 1, \ldots, m$

where j1 and j2 denote the negative and positive criteria, respectively.

The following table shows both positive and negative ideal values.

|  | Positive ideal | Negative ideal |
|---|---|---|
| Sustainability | 0.12 | 0.074 |
| Agility | 0.128 | 0.051 |
| Cost Optimization | 0.123 | 0.061 |
| Visibility | 0.117 | 0.081 |
| Technological integration | 0.134 | 0.046 |

**Table 7.** The positive and negative ideal values

**STEP4: distance from the positive and negative ideal solutions**

TOPSIS method ranks each alternative based on the relative closeness degree to the positive ideal and distance from the negative ideal. Therefore, in this step, the calculation of the distances between each alternative and the positive and negative ideal solutions is obtained by using the following formulas.

$$d_i^+ = \sqrt{\sum_{j=1}^{n}[v_{ij}(x) - v_j^+(x)]^2} \quad , \quad i = 1, \dots, m$$

$$d_i^- = \sqrt{\sum_{j=1}^{n}[v_{ij}(x) - v_j^-(x)]^2} \quad , \quad i = 1, \dots, m$$

The following table shows the distance to the positive and negative ideal solutions.

|  | Distance to positive ideal | Distance to negative ideal |
|---|---|---|
| Technological Proficiency | 0.037 | 0.137 |
| Collaborative Ecosystem | 0.134 | 0.043 |
| Agility and Responsiveness | 0.059 | 0.109 |
| Cost Efficiency and Optimization | 0.085 | 0.08 |

**Table 8.** Distance to positive and negative ideal points

**STEP 5: Calculate the relative closeness degree of alternatives to the ideal solution**

In this step, the relative closeness degree of each alternative to the ideal solution is obtained by the following formula. If the relative closeness degree has value near to 1, it means that the alternative has shorter distance from the positive ideal solution and longer distance from the negative ideal solution.

$$C_i = \frac{d_i^-}{(d_i^+ + d_i^-)} \quad , \quad i = 1, \dots, m$$

The following table shows the relative closeness degree of each alternative to the ideal solution and its ranking.

|  | Ci | rank |
|---|---|---|
| Technological Proficiency | 0.787 | 1 |
| Collaborative Ecosystem | 0.242 | 4 |
| Agility and Responsiveness | 0.651 | 2 |
| Cost Efficiency and Optimization | 0.484 | 3 |

**Table 9.** Ci values and ranking

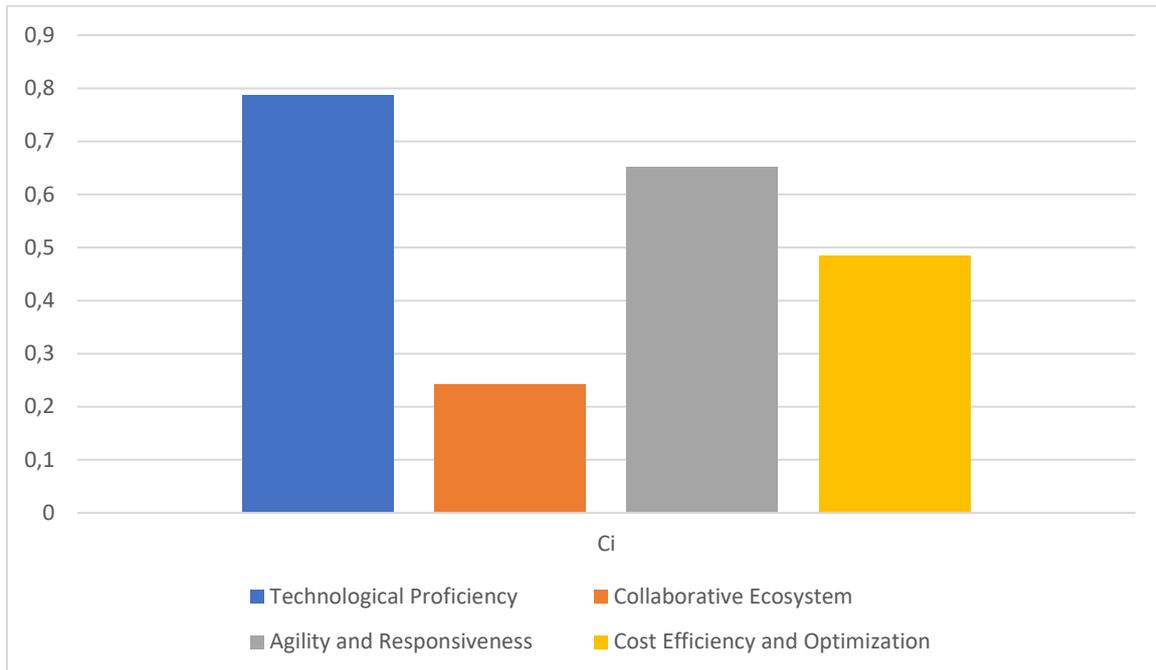

**Figure 2:** Maturity Factor Ranking

Through the Grey-TOPSIS analysis, four relevant maturity factors were revealed that affected the enhancement of supply chain resilience in the following order of priority: technological proficiency as the most influential, agility and responsiveness next, cost efficiency and optimization third, and collaborative ecosystem fourth. These results will allow us to gain useful information on the relevance of each factor and its direct influence toward achieving robust and responsive supply chains less susceptible to disruption and uncertainties in the marketplace.

**Technology Proficiency** ranks the highest with the closeness index ($C_1 = 0.787$), thus confirming the fact that it is the most important in strengthening the resilience of supply chain. It points out that organizations which invest in innovative technologies like AI, blockchain, IoT, and automation are more capable of tracking supply chain processes in real time and predicting and preemptively attacking potential disruptions. Technology integration provides a level of visibility, predictive possibilities, and operational efficiencies that allow the business to see the risks earlier and adjust more rapidly. And that, put companies in a much stronger position when the supply chain disruptions are being able to leverage data-driven insights to improve logistics, improve stocking, improve purchasing, and so on. This outcome indicates that in the digital age, technological progress is a need, not a luxury, for resilience and lasting sustainability.

The second is Agility and Responsiveness ($C_i = 0.651$), which signifies that flexible and adaptable supply chain movements are important as well. Being able to react swiftly according to ever-changing market demands, supply shortages, or non-food/political redirection is vital in minimizing possible disruptions and ensuring the continuity of business. Then, agile supply chains are able to change sourcing, redirect shipments, scale up or down production all very quickly, thereby minimizing the blow of an external shock. This outcome reflects how resilience is now at the heart of an organization's ability to stay dynamic and adaptable. Companies that embrace more agile processes and structures for decision-making will be better able to resume normal operations than companies with less responsive structures.

**Cost Efficiency and Optimization** ($C_i = 0.484$) is ranked third, showing that although costs and supply chain optimization are important factors of supply chain management, they still follow technological and operational flexibility in affecting resilience. Emphasis on lean and eliminating surplus costs can help profitability and waste, but too much emphasis on COST EFFICIENCY without flexibility and redundancy can create problems. These findings imply that cost optimization or cost minimization has only a supporting influence on resilience and that is should not be cultivated to the point that is retards the responsiveness and adaptability of the supply chain. Companies that find a way to balance the two achieve high inventory turnover while still being resilient to supply chain disruptions.

**Collaborative Community** ranked the lowest ($C_i = 0.242$) only to emphasize the comparative significance of collaboration and interorganizational partnerships in building resilience. Collaboration with suppliers, partners, and stakeholders is required for joint risk management, information sharing, and response coordination, but its figure one ranking indicates that collaboration alone does not achieve the immediate, groundbreaking results that technology and agility can. Which could simply be indicative of the fact that unless one has top-notch technical skills and nimble-station architectures, it's only possible to produce so much through collaboration. However, continuing the creation of a cooperative environment is still an important element of long-term resilience in that it fosters connections, promotes trust, and strengthens the community's ability to endure changes.

In conclusion, the Grey-TOPSIS findings highlight that technological progress and operational flexibility are the primary factors of supply chain resilience, while cost effectiveness and collaboration serve as pivotal yet supportive elements. Entities aiming to boost their resilience must focus on investing in digital transformation, automation, and agile methods to develop strong and flexible supply chains. Although cost optimization and collaborative ecosystems enhance resilience, they need to be combined with technological and agile initiatives to form a comprehensive, resilient supply chain that can effectively handle future uncertainties and maintain a competitive edge.

These rankings reflect the role of each maturity factor in building supply chain resilience, directing organizations on strategic priorities to improve their operational efficiency in a more complex landscape.

# 5. Discussion

The outcomes of the analyses employing the AHP methodology and Grey Topsis offer important perspectives on the elements affecting supply chain resilience, especially regarding the improvement of operations via technologies such as the metaverse and ChatGPT. The AHP approach recognized five essential criteria—Sustainability, Agility, Cost Optimization, Visibility, and Technological Integration—as the key factors for enhancing supply chain resilience. This emphasis highlights the necessity for organizations to concentrate on these sectors to create strong and flexible supply chains capable of enduring disruptions.

Technological proficiency is becoming a vital component of this capability as it endows organizations with more enhanced tools and creative platforms. Among these, the Metaverse and AI-enabled applications like ChatGPT are transforming the way businesses improve visibility, agility, and interaction throughout their supply chains.

The metaverse—a fully immersive, three-dimensional digital space—promises to be the ultimate space for supply chain efficiency. With simulated scenarios and on-demand monitoring, companies recreate a virtual twin of their supply chain, which provides accurate viewing and diagnostic of their operations. This ability improves visibility to track inventory, logistics, and bottlenecks - before they become problems. Virtual environments, for example, allow stakeholders to collaborate remotely and, despite their distance, can make efficient and well-informed decisions across the span of continents.

Moreover, the metaverse supports agility and responsiveness by allowing businesses to simulate various scenarios, such as supplier disruptions or demand spikes. These simulations provide actionable insights into the potential impact of such events, helping organizations prepare contingency plans and optimize resource allocation. The integration of metaverse technologies into supply chain operations fosters resilience by enhancing decision-making and reducing the response time to disruptions. (Muhammad Turki Alshurideh, Barween Al Kurdi, Yousef ahmad Damra, Sara Yasin 2024)

Tools such as AI (ChatGPT) help supplement the resilience of the supply chain. And with ChatGPT ingesting oceans of data and understanding their implications, it can discern patterns, predict disruptions, and deliver actionable insights continuously. This function facilitates technology in providing a means to link communication gaps between different facets of the supply chain. The Effects of ChatGPT on Supply Chain Efficiency and Agility, 2023.

For instance, ChatGPT could act as a smart assistant to a supply chain manager, answering any question on demand, such as inventory levels, delivery delays, or supplier performance. Due to its human-like text generation capability, collaboration in the supply chain is supported as everyone on the team can collaborate to solve a problem more efficiently. Furthermore, by automating routine tasks like demand forecasting, order processing, etc., ChatGPT can free up resources for more strategic initiatives. (Zycus, 2024)

Being technologically savvy also helps promote sustainability and prevent risk, two key components of a resilient supply chain. By using the metaverse to find out exactly how they are either harming or helping the environment, organizations will have the visibility of the different footprints and hopefully use this type of technology to switch to more green methods of life. ChatGPT's data-driven insights similarly enable managers to foresee potential threats like vulnerabilities in suppliers or geopolitical disturbances, which then facilitates the implementation of preventative strategies. (Guilherme Francisco Frederico, 2023)

Integrating the metaverse and ChatGPT bolsters supply chain resiliency through a synergy, which can be considered a resilient final interpretation. By providing immersive visualization and intelligent decision-support for very complex operations, they manage both. This partnership converts conventional supply chains into flexible, vibrant networks that can grow in an uncertain business environment. (Xiang Huang and Ping-Kuo Chen, 2023).

In summary, technological proficiency, in the form of the metaverse and ChatGPT, is transforming the world to a place with more resilient supply chains. With these new tools, organizations will be able to increase supply chain visibility, agility, sustainability, and risk awareness to protect their supply chains from disruptions. Technology is constantly changing at an exponential pace, and these kinds of advances will be necessary to remain successful in the increasingly complicated global market of the future.

## 6. Conclusion

Integrating Metaverse technologies and ChatGPT into supply chain management could transform and strengthen resilience substantially. In our study, we applied the Analytic Hierarchy Process (AHP) and G-TOPSIS methods and determined the five primary criteria that most enhance resilient supply chains: Sustainability, Agility, Cost Optimization, Visibility, and Technological Integration. Technology Competence was one of these major maturity factors required to best utilize these capabilities.

In summary, companies looking to improve their supply chain resilience should focus on the following as strategic criteria. By focusing on sustainability, they can ensure long-term viability and adaptability in a rapidly changing environment. Agility gives the ability to move quickly in response to disruptions and cost optimization protects the bottom line in difficult times. Visibility throughout the supply chain increases transparency and allows for more intelligent decision-making, and the implementation of technology creates a cohesive interaction of information and activities. Also, creating technological proficiency is an important factor here. With technology playing a more pervasive role in supply chains through artificial intelligence, data analytics, and other methods, companies need to develop their technological ability to fully exploit these modern technologies. Such competence will enhance performance as well as lead to the development of innovative ideas.

In conclusion, if companies successfully embrace these results and act on these suggestions, they can drastically improve the reliability of their supply chains and be poised to succeed in the uncertain global environment. Supply chains are the future of technology, collaboration, and adaptation in an ever-expanding world.

# Appendix 1 : AHP Calculation for expert 4

Table A1 shows the assessment of capabilities using a scale from 1 to 9.

| Capabilities | Visibility | Agility | Collaboration | Flexibility | Security | Cost Optimization | Technological integration | Risk Prevention | Sustinability | Organizational Capacity |
|---|---|---|---|---|---|---|---|---|---|---|
| Visibility | 1 | 3 | 1 | 1 | 0,5 | 0,2 | 2 | 0,5 | 0,333333 | 1 |
| Agility | 0,33333 | 1 | 1 | 1 | 0,2 | 0,333333 | 1 | 0,333333 | 1 | 0,333333 |
| Collaboration | 1 | 1 | 1 | 1 | 0,3333 | 0,142857 | 1 | 0,2 | 0,333333 | 1 |
| Flexibility | 1 | 1 | 1 | 1 | 0,3333 | 0,2 | 1 | 0,25 | 0,2 | 1 |
| Security | 2 | 5 | 3 | 3 | 1 | 0,2 | 1 | 0,333333 | 1 | 3 |
| Cost Optimization | 5 | 3 | 7 | 5 | 5 | 1 | 5 | 5 | 5 | 5 |
| Technological integration | 0,5 | 1 | 1 | 1 | 1 | 0,2 | 1 | 1 | 1 | 3 |
| Risk Prevention Awareness | 2 | 3 | 5 | 4 | 3 | 0,2 | 1 | 1 | 3 | 3 |
| Sustinability | 3 | 1 | 3 | 5 | 1 | 0,2 | 1 | 0,333333 | 1 | 3 |
| Organizational Capacity | 1 | 3 | 1 | 1 | 0,3333 | 0,2 | 0,333333 | 0,333333 | 0,333333 | 1 |
| Sum | 16,8333 | 22 | 24 | 23 | 12,7 | 2,87619 | 14,333333 | 9,283332 | 13,199999 | 21,333333 |

**Table A1. Pair-wise comparison matrix**

The normalized pairwise comparison matrix (Table A2) is created by dividing each element in a row by the sum of its respective column. The weights are then determined by calculating the average of all elements within each row.

| Capabilities | Visibility | Agility | Collaboration | Flexibility | Security | Cost Optimiza | Technological | evention Awa | Sustinability | Organizational |
|---|---|---|---|---|---|---|---|---|---|---|
| Visibility | 0,05941 | 0,136 | 0,041666667 | 0,04348 | 0,0394 | 0,069536435 | 0,139534887 | 0,05385997 | 0,0252525 | 0,046875001 |
| Agility | 0,0198 | 0,045 | 0,041666667 | 0,04348 | 0,0157 | 0,115893943 | 0,069767443 | 0,03590661 | 0,07575758 | 0,015624985 |
| Collaboration | 0,05941 | 0,045 | 0,041666667 | 0,04348 | 0,0262 | 0,049668833 | 0,069767443 | 0,02154399 | 0,0252525 | 0,046875001 |
| Flexibility | 0,05941 | 0,045 | 0,041666667 | 0,04348 | 0,0262 | 0,069536435 | 0,069767443 | 0,02692999 | 0,01515152 | 0,046875001 |
| Security | 0,11881 | 0,227 | 0,125 | 0,13043 | 0,0787 | 0,069536435 | 0,069767443 | 0,03590661 | 0,07575758 | 0,140625002 |
| Cost Optimizatio | 0,29703 | 0,136 | 0,291666667 | 0,21739 | 0,3937 | 0,347682177 | 0,348837217 | 0,53859972 | 0,37878791 | 0,234375004 |
| Technological int | 0,0297 | 0,045 | 0,041666667 | 0,04348 | 0,0787 | 0,069536435 | 0,069767443 | 0,10771994 | 0,07575758 | 0,140625002 |
| Risk Prevention | 0,11881 | 0,136 | 0,208333333 | 0,17391 | 0,2362 | 0,069536435 | 0,069767443 | 0,10771994 | 0,22727274 | 0,140625002 |
| Sustinability | 0,17822 | 0,045 | 0,125 | 0,21739 | 0,0787 | 0,069536435 | 0,069767443 | 0,03590661 | 0,07575758 | 0,140625002 |
| Organizational C | 0,05941 | 0,136 | 0,041666667 | 0,04348 | 0,0262 | 0,069536435 | 0,023255791 | 0,03590661 | 0,0252525 | 0,046875001 |

**Table A2. Normalized pairwise comparison matrix**

The priority vector is a crucial component that represents the relative importance of different criteria or alternatives based on pairwise comparisons. It is derived from a pairwise comparison matrix, where decision-makers evaluate the importance of each criterion relative to others.

It's calculated by defining the average of each row of Table A2.

Matrix multiplication plays a vital role in synthesizing judgments and deriving priority vectors from pairwise comparison matrices. Here's an overview of how matrix multiplication is applied within AHP. It represents the Matrix product of the Pairwise Comparisons Matrix and the Priority Vector.

The weight is calculated as following:

## Weight= The priority vector/ Matrix multiplication

| PRIORITY VECTOR | Matrice Multiplications | WEIGHTED SUM MATRIX/PV |
|---|---|---|
| 0,065534338 | 0,714191399 | 10,89797221 |
| 0,047910003 | 0,525159472 | 10,9613742 |
| 0,042935988 | 0,467409608 | 10,8861967 |
| 0,044451249 | 0,479230698 | 10,78104005 |
| 0,107185263 | 1,179545217 | 11,00473313 |
| 0,318443416 | 3,716253 | 11,67005758 |
| 0,070244901 | 0,8140696 | 11,58902047 |
| 0,148856396 | 1,7349316 | 11,65506925 |
| 0,103639691 | 1,142341217 | 11,02223677 |
| 0,050798754 | 0,554444457 | 10,91452863 |

## Table A3. Weight SUM Matrix Calculation

The consistency ratio is calculated as following :

**CR = CI/RI.**

With

$$CI = \frac{\lambda_1 \max - n}{n - 1}$$

And $\lambda_1 max$ is the average of the Weight Sum Calculated in the Table A3.

| $\lambda_{max}$ | 11,1382229 |
|---|---|
| CI | 0,126469211 |
| CR | 0,084878665 |

## Table A4. Consistency Ratio

**Reference:**

1. *Sylvie Michel, Sylvie Gerbaix, Bernard Marc, 2023.*
2. *Maureen S. Golan, Laura H. Jernegan, Igor Linkov, 2020.*
3. *Weili Yin, Wenxue Ran, 2022.*
4. *Xianjun Zhu, Yenchun Jim Wu, 2022.*
5. *Youan Ke, Lin Lu, Xiaochun Luo, 2023.*
6. *Benjamin R. Tukamuhabwa, Mark Stevenson, Jerry Busby, Marta Zorzini, 2015.*
7. *Andreas Wieland, Christian F. Durach, 2021.*
8. *Timothy J. Pettit Ph.D, Joseph Fiksel Ph.D., Keely L. Croxton Ph.D., MIT, 2011.*
9. *Carla Roberta Pereira, Martin Christopher, Andrea Lago Da Silva, 2014.*
10. *Kirstin Scholten, Sanne Schilder, 2015.*
11. *Asmae El Jaouhari, Jabir Arif, Fouad Jawab, Ashutosh Samadhiya, Anil Kumar, 2024.*
12. *Ping-Kuo Chen, Xiang Huang, 2023 https://doi.org/10.1002/sd.2663*
13. *Dmitry Ivanov, Jennifer Blackhurst, Ajay Das, 2021.*
14. *ChatGPT in Supply Chains: Initial Evidence of Applications and Potential Research Agenda, Guilherme Francisco Frederico, 2023*
15. *Samuel Fosso Wamba , Maciel M. Queiroz ,Charbel Jose Chiappetta Jabbour , Chunming (Victor) Shi, 2023.*
16. *Mohd Javaid , Abid Haleem , Ravi Pratap Singh, 2023.*
17. *Abid Haleem , Mohd Javaid , Ravi Pratap Singh, 2022.*
18. *Stephen Rice , Sean R. Crouse, Scott R. Winter , Connor Rice, 2024.*
19. *Xinyue Chu, Yizhou Wang, Qi Chen, Jiaquan Gao, 2022.*
20. *Hainan Ye, Lei Wang, 2023.*
21. *Omar Ali, Peter A. Murray, Mujtaba Momin, Fawaz S. Al-Anzi, 2023.*
22. *Igor Calzada, 2023*
23. *Shiying Zhang, Jun Li, Long Shi, Ming Ding, Dinh C. Nguyen, Wen Chen, Zhu Han, 2024.*
24. *Alexandre Dolgui, Dmitry Ivanov, 2023*
25. *Rohit Raj, Arpit Singh, Vimal Kumar, Pratima Verma, 2023.*
26. *Yamin Ma, 2023.*
27. *Yiheng Liu, Tianle Han, Siyuan Ma, Jiayue Zhang, Yuanyuan Yang, Jiaming Tian, Hao He a, Antong Li, Mengshen He, Zhengliang Liu, Zihao Wu, Lin Zhao, Dajiang Zhu, Xiang Li, Ning Qiang, Dingang Shen, Tianming Liu, Bao Ge, 2023.*
28. *Partha Pratim Ray, 2023.*
29. *Yatao Li,  Jianfeng Zhan, 2022.*
30. *Mohd Javaid, Abid Haleem, Ravi Pratap Singh, Shahbaz Khan, Ibrahim Haleem Khan, 2023.*
31. *Vargas, R. V. (2010). Using the analytic hierarchy process (ahp) to select and prioritize projects in a portfolio. Paper presented at PMI® Global Congress 2010—North America, Washington, DC. Newtown Square, PA: Project Management Institute.https://www.pmi.org/learning/library/analytic-hierarchy-process-prioritize-projects-6608?*
32. *Khadija Echefaj, Abdelkabir Charkaoui, Anass Cherrafi, Anil Kumar, Sunil Luthra 2022*
33. *Journal of Global Operations and Strategic Sourcinghttps://www.emerald.com/insight/content/doi/10.1108/jgoss-05-2022-0040/full/html?skipTracking=true*
34. *Unlocking the Power of Supply Chain Resiliency with Industrial Metaverse, Oyku_Ilgar 2023 https://community.sap.com/t5/supply-chain-management-blogs-by-sap/unlocking-the-power-of-supply-chain-resiliency-with-industrial-metaverse/ba-p/13579920?utm_source=chatgpt.com*

35. *Leveraging Industry 4.0 Technologies for Sustainable Humanitarian Supply Chains: Evidence from the Extant Literature by M. Ali Ülkü ,ORCID,James H. Bookbinder,and Nam Yi Yun https://doi.org/10.3390/su16031321*
36. *Resilience and cleaner production in industry 4.0: Role of Supply Chain Mapping and Visibility, MUBARIK, Muhammad Shujaat, NAGHAVI, Navaz, MUBARIK, Mobashar, http://shura.shu.ac.uk/32469/*
37. *supply chain mapping and visibility*
38. *Enhancing Supply Chain Resilience Through Artificial Intelligence: Developing a Comprehensive Conceptual Framework for AI Implementation and Supply Chain Optimization y Meriem Riad 1, Mohamed Naimi, and Chafik Okar. https://doi.org/10.3390/logistics8040111*
39. *Multi-Criteria Decision Making (MCDM) Methods and Concepts, Hamed Taherdoost and Mitra Madanchian 2023 https://doi.org/10.3390/encyclopedia3010006*
40. *Multi-Criteria Decision-Making (MCDM) as a powerful tool for sustainable development: Effective applications of AHP, FAHP, TOPSIS, ELECTRE, and VIKOR in sustainability Nitin Rane, Anand Achari, and Saurabh Purushottam Choudhary 2023 http://dx.doi.org/10.56726/IRJMETS36215*
41. *Hamed Taherdoost. Decision Making Using the Analytic Hierarchy Process (AHP); A Step by Step Approach. International Journal of Economics and Management System, 2017. ffhal-02557320f https://hal.science/hal-02557320v1*
42. *Jalaliyoon, N., Bakar, N. A., Taherdoost, H. (2012). Accomplishment of Critical Success Factor in Organization; Using Analytic Hierarchy Process. International Journal of Academic Research in Management, Helvetic Editions Ltd, 1(1); 1-9.*
43. *Golden, B. L. & Wang, Q. (1990). An Alternative Measure of Consistency. In: B. L. Golden, A. Wasil & P.T. Harker (eds.) Analytic Hierarchy Process: Applications and*
44. *] Lee. M. C. (2007). A Method of Performance Evaluation by Using the Analytic Network Process and Balanced Score Card, International Conference on Convergence Information Technology.*
45. *Saaty, T. L. (1980). The Analytic Hierarchy Process: Planning, Priority Setting, Resources Allocation. London: McGraw-Hill.*
46. *A Modified TOPSIS Method utilizing the Gray Correlation Analysis, Guan-Dao Yang; Lu Sun; Xiao Liu, 2010 10.1109/i-Society16502.2010.6018767*
47. *Application of Grey-TOPSIS approach to evaluate value chain performance of tea processing chains, Richard Nyaogaa\*, Peterson Magutub and Mingzheng Wangc, 2016 https://www.growingscience.com/dsl/Vol5/dsl_2016_2.pdf?utm_source=chatgpt.com*
48. *Application of Grey TOPSIS in Preference Ordering of Action Plans in Balanced Scorecard and Strategy Map, Mohammadreza SADEGHI1∗, Seyed Hossein RAZAVI1, Narges SABERI2 2012 INFORMATICA, 2013, Vol. 24, No. 4, 619–635*
49. The impact of the metaverse on the Supply Chain, Dave Food 2022
50. *Exploring the impact of metaverse adoption on supply chain effectiveness: A pathway to competitive advantage, Muhammad Turki Alshurideh, Barween Al Kurdi, Yousef ahmad Damra, Sara Yasin, 2024, 10.5267/j.uscm.2023.12.017*
51. *The Impact of ChatGPT on Supply Chain Efficiency and Agility, 2023*
52. *Harnessing ChatGPT for Supply Chain Efficiency, Zycus, 2024*
53. *A systematic literature review exploring the relationship between metaverse and supply chain resilience: the role of sensory feedback Xiang Huang, Ping-Kuo Chen https://doi.org/10.1145/3589860.3589872*
54. *COMPARISON OF AHP AND ANP METHODS FOR RESILIENCE MEASUREMENT IN SUPPLY CHAINS, Pavel WICHER, Radim LENORT, 2014*

*https://www.confer.cz/metal/2014/download/3200-comparison-of-ahp-and-anp-methods-for-resilience-measurement-in-supply-chains.pdf*
55. *Developing Supply Chain Maturity, ADRIAN DONE, 2011 https://www.iese.edu/media/research/pdfs/DI-0898-E.pdf*